\documentclass[%
 aip,rsi,amsmath,amssymb,reprint,
]{revtex4-1}

\usepackage{graphicx}
\usepackage{dcolumn}
\usepackage{bm}
\usepackage{lipsum}
\usepackage{xcolor}

\begin{document}

\preprint{AIP/123-QED}

\title{Quantum reservoir computing implementation on coherently coupled quantum oscillators}

\author{Julien Dudas}
\author{Baptiste Carles}
\author{Erwan Plouet}
\author{Alice Mizrahi}
\author{Julie Grollier}
\author{Danijela Markovi\'c}\thanks{Author to whom correspondence should be addressed. Electronic mail : danijela.markovic@cnrs-thales.fr}

 \affiliation{Unit\'e Mixte de Physique CNRS, Thales, Universit\'e Paris-Saclay, 91767 Palaiseau, France}

\begin{abstract}
Quantum reservoir computing is a promising approach for quantum neural networks, capable of solving hard learning tasks on both classical and quantum input data. However, current approaches with qubits suffer from limited connectivity. We propose an implementation for quantum reservoir that obtains a large number of densely connected neurons by using parametrically coupled quantum oscillators instead of physically coupled qubits. We analyse a specific hardware implementation based on superconducting circuits: with just two coupled quantum oscillators, we create a quantum reservoir comprising up to 81 neurons. We obtain state-of-the-art accuracy of 99 \% on benchmark tasks that otherwise require at least 24 classical oscillators to be solved.
Our results give the coupling and dissipation requirements in the system and show how they affect the performance of the quantum reservoir. Beyond quantum reservoir computing, the use of parametrically coupled bosonic modes holds promise for realizing large quantum neural network architectures, with billions of neurons implemented with only 10 coupled quantum oscillators.
\end{abstract}

\maketitle

\section*{Introduction}

Quantum neural networks are the subject of intensive research today. They emulate a large number of neurons with only a small number of physical components, which facilitates scaling up compared to classical approaches. Indeed, by encoding the responses of the neurons in the populations of basis states, a system of $N$ qubits provides up to $2^N$ neurons. Moreover, such quantum neural networks could automatically transform complex quantum data into simple outputs representing the class of the input, that could then be measured with just a few samples compared to the millions needed today. The first experimentally realized quantum neural networks are made of 7~\cite{Herrmann2021} to 40 qubits~\cite{Huang2022}, each connected to two or four nearest neighbors. 

The major challenge of the field is now to experimentally realize neural networks capable of real-world classification tasks, and thus containing millions of neurons each connected by thousands of connections. The use of qubits poses a conceptual and technical problem for this purpose. Indeed, when connectivity is obtained with pairwise couplers between qubits, distant qubits cannot be interconnected without very cumbersome classical circuitry in existing 2D architectures. 

Here we develop an alternative approach to quantum neural networks that is both scalable and compatible with experimental implementations. We propose to leverage the complex dynamics of coherently coupled quantum oscillators, combined with their infinite number of basis states, to populate a large number of neurons much more efficiently than with qubits \cite{Dudas2022}. We then show through simulations with experimentally-validated models that this system classifies and predicts time-series data with high efficiency through the approach of reservoir computing~\cite{Haas2004}.  

With just two quantum oscillators, up to 9 states can be populated in each oscillator with significant probability amplitudes, which yields a quantum reservoir with up to 81 neurons.

We evaluate its performance on two benchmark tasks in the field of reservoir computing, sine-square waveform classification and Mackey-Glass chaotic time series prediction, that test the ability of the reservoir to memorize input data and transform it in a non-linear way. We obtain a state-of-the-art accuracy of 99 \% with our system of two coherently coupled oscillators, which otherwise requires 24 classical oscillators to achieve. With 10 oscillators we could have 3 billions of neurons, comparable to the most impressive neural networks capable of hard tasks such as natural language processing or generating images from text descriptions~\cite{Ramesh2022}. Furthermore, these neurons can be populated in a much more efficient way compared to those implemented on qubit systems, by resonantly driving each of the oscillators and coupling them strongly pairwise.

The results show that two coupled quantum oscillators implement a high quality reservoir computer capable of complex tasks, and open the path to experimental implementations of quantum reservoirs based on a large number of basis state neurons, thus providing a quantum neural network platform compatible with numerous algorithms exploiting physics and dynamics for computing~\cite{Amin2018, Scellier2017, Holder2019, Onodera2020}. 

\section*{Quantum reservoir computing}

Reservoir computing is a machine learning paradigm that uses nonlinear dynamical systems for temporal information processing~\cite{Haas2004}. Its principle is illustrated in FIG.~1. The reservoir (blue area) is a dynamical system with arbitrary but fixed recurrent connections. It takes as input data that is not easily separable in different classes. The role of the reservoir is to project these inputs into a highly dimensional state space in which the data becomes linearly separable. The reservoir outputs are then classified by a linear, fully connected layer (shown in red arrows) that can be trained by a simple linear regression. Physical implementations of reservoir computing commonly perform this projection of input data to a high dimensional space through complex non-linear dynamics and the outputs are obtained by measuring specific variables on that system~\cite{Paquot2012, Brunner2013, Vandoorne2014, Torrejon2017}.

The fully connected layer is usually realized in software, and multiplies the measurement outputs $F(X)$ by a weight matrix $W$, such that
\begin{equation}
    W F(X) = Y.
\end{equation}
The weight matrix is trained to make the neural network output $Y$ match the target vector $\tilde{Y}$. The particularity of reservoir computing compared to deep neural networks is that training can be performed in a single step by matrix inversion
\begin{equation}
    W = {\tilde{Y}_{\rm{train}}} F^\dagger(X_{\rm{train}}),
\end{equation}
where $X_{\rm{train}}$ is the training data, $\tilde{Y}_{\rm{train}}$ is the training target, and $F^\dagger$ is the Moore-Penrose pseudo-inverse of the matrix $F$ containing the outputs $f(x_i)$ of the reservoir neurons for all the training examples~\cite{Appeltant2011, Brunner2013}. The matrix inversion method works well for small matrix dimensions. For larger matrix dimensions other methods can be used, such as linear regression used in reference~\cite{Rafayelyan2020} for a 50 000 node classical reservoir.
The learned weight matrix is applied on the test data contained in the vector $X_{\rm{test}}$, in order to find the neural network prediction 
\begin{equation}
    Y_{\rm{test}} = W F(X_{\rm{test}}).
    \label{prediction}
\end{equation}
Comparing the prediction to the test target $\tilde{Y}_{\rm{test}}$ allows to evaluate the prediction accuracy, i.e. the fraction of times the data point is correctly classified, as well as the normalized root mean square error
\begin{equation}
   \rm{NRMSE} = \frac{1}{y_{\rm{max}} - y_{\rm{min}}} \sqrt{\frac{\sum_i^N (y_i - \tilde{y_i})^2}{N}}.
\end{equation}
 We will use these measures to evaluate the performance of the reservoir, and thus the capacity of the chosen dynamical system to efficiently implement reservoir computing. It should be noted that there are neural networks capable of achieving higher accuracies than reservoir computing for the learning tasks investigated in this study~\cite{Prater2017a}. Our objective is not to contrast various algorithms, but rather, for a specific task, employing a straightforward network architecture and a designated training method, to assess the benefits conferred by distinct physical properties associated with diverse physical platforms, including quantum aspects, in enhancing the system's computing capabilities.

\begin{figure}
\includegraphics[scale=0.2]{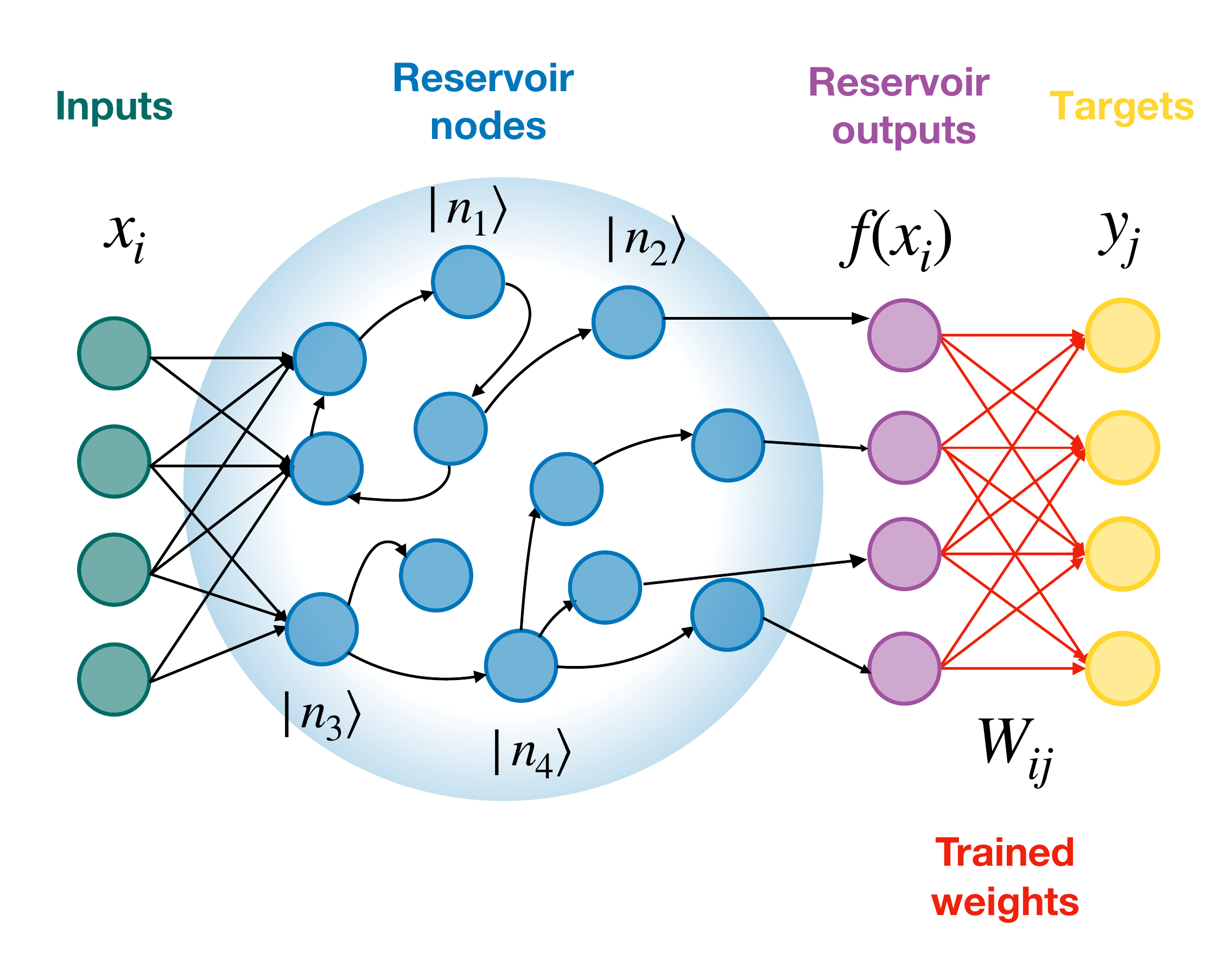}
\caption{Principle of quantum reservoir computing. The reservoir neurons (blue circles) are basis states of a coupled quantum system. The reservoir outputs (purple circles) are the measured occupation probabilities of these states. The black connections in the physical reservoir (blue area) are fixed and the red connections are trained.}
\label{quant_res}
\end{figure}

 Reservoir computing was implemented on different classical physical systems, ranging from silicon photonics~\cite{Vandoorne2014} and optoelectronics~\cite{Paquot2012, Brunner2013}, to spintronic nano-oscillators~\cite{Torrejon2017}.
Quantum reservoir computing was first proposed in 2017, with a reservoir whose neurons correspond to the basis states of a set of qubits, and computational capabilities are identical to 100-500 classical neurons with only 5-7 qubits~\cite{Fujii2017}. Experimentally, quantum reservoir was implemented with 4 static spins and 8 neurons \cite{Negoro2018}, and a dissipative reservoir was implemented with up to 10 qubits~\cite{Chen2020a}. We have recently highlighted that dynamical systems of coherently coupled quantum oscillators possess all the required features for quantum reservoir computing\cite{Dudas2022}.

\section*{Coupled quantum oscillators}

\begin{figure}
\includegraphics[scale=0.3]{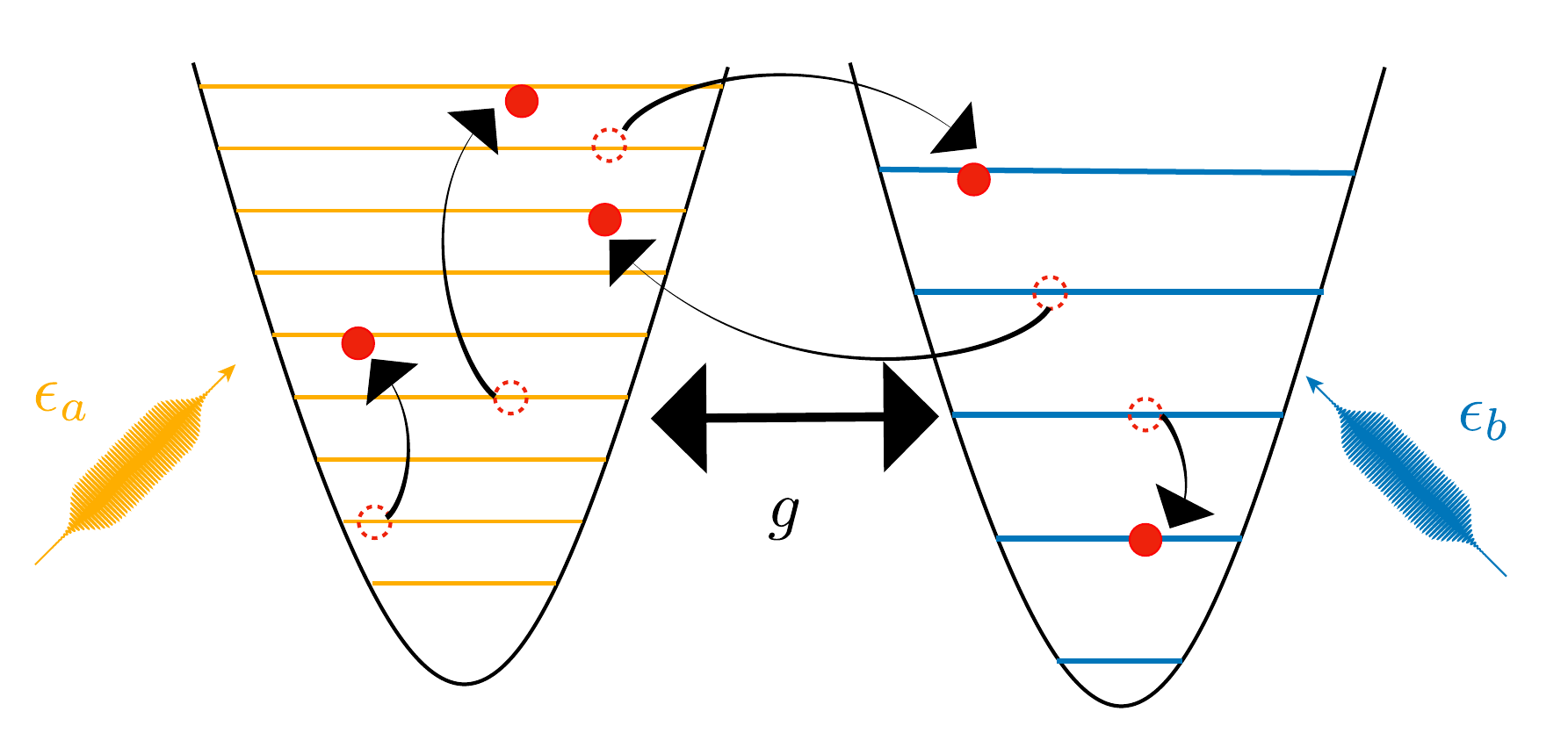}
\caption{Schematic of two quantum oscillators with different resonance frequencies $\omega_a$ and $\omega_b$ and energy levels separated by $\hbar \omega_a$ and $\hbar \omega_b$. The oscillators are resonantly driven at amplitudes $\epsilon_a$ and $\epsilon_b$, with dissipation rates $\kappa_a$ and $\kappa_b$. The coherent coupling at a rate $g$ results in the exchange of excitations between oscillators.}
\label{JPC}
\end{figure}

We consider the implementation of quantum reservoir computing with two coupled quantum oscillators $a$ and $b$ (FIG.~\ref{JPC}). In this case, the reservoir neurons are given by the basis states $| n_a, n_b \rangle$, and the reservoir outputs by their occupation probabilities. Such a system can be experimentally implemented using superconducting circuits featuring resonators whose fundamental modes are coupled using three- or four-wave mixing elements such as Josephson mixers\cite{Bergeal2010, Abdo2013, Abdo2014}, SNAIL\cite{ N.E.Frattini2017} or even a single transmon\cite{ N.E.Frattini2017, Gao2018}. The Hamiltonian describing such system writes~\cite{Abdo2013a}
\begin{equation}
{\hat{H}}= \hbar \omega_a\hat{a}^{\dagger}\hat{a} + \hbar \omega_b\hat{b}^{\dagger}\hat{b} + g\left(\hat{a}{\hat{b}}^\dag+{\hat{a}}^\dag\hat{b}\right),
\end{equation}
where $g=\chi p$ is the parametric conversion coupling rate that can typically be controlled by the amplitude of a pump tone. Such tunable coupling allows us to study the performance of the quantum reservoir as a function of the coupling strength.

We drive each oscillator at resonance with an amplitude that encodes the input data, such that the population of the basis states depends on this input value, the duration for which the drive signal is applied, and the previous input values, as long as each input is sent for a time shorter than the lifetime  of the oscillators. The dynamics of the system is driven by three main processes: resonant drives, dissipation and conversion of photons between the oscillators at a rate $g$.  We use typical experimental parameters for superconducting circuits with resonators frequencies
$\omega_a = 2 \pi \times 10$ GHz and $\omega_b = 2 \pi \times 9$ GHz and dissipation rates $\kappa_a = 2 \pi \times 17$ MHz, $\kappa_b = 2 \pi \times 21$ MHz. 

\section*{Learning tasks with the quantum reservoir} 

\begin{figure}
\includegraphics[scale=0.4]{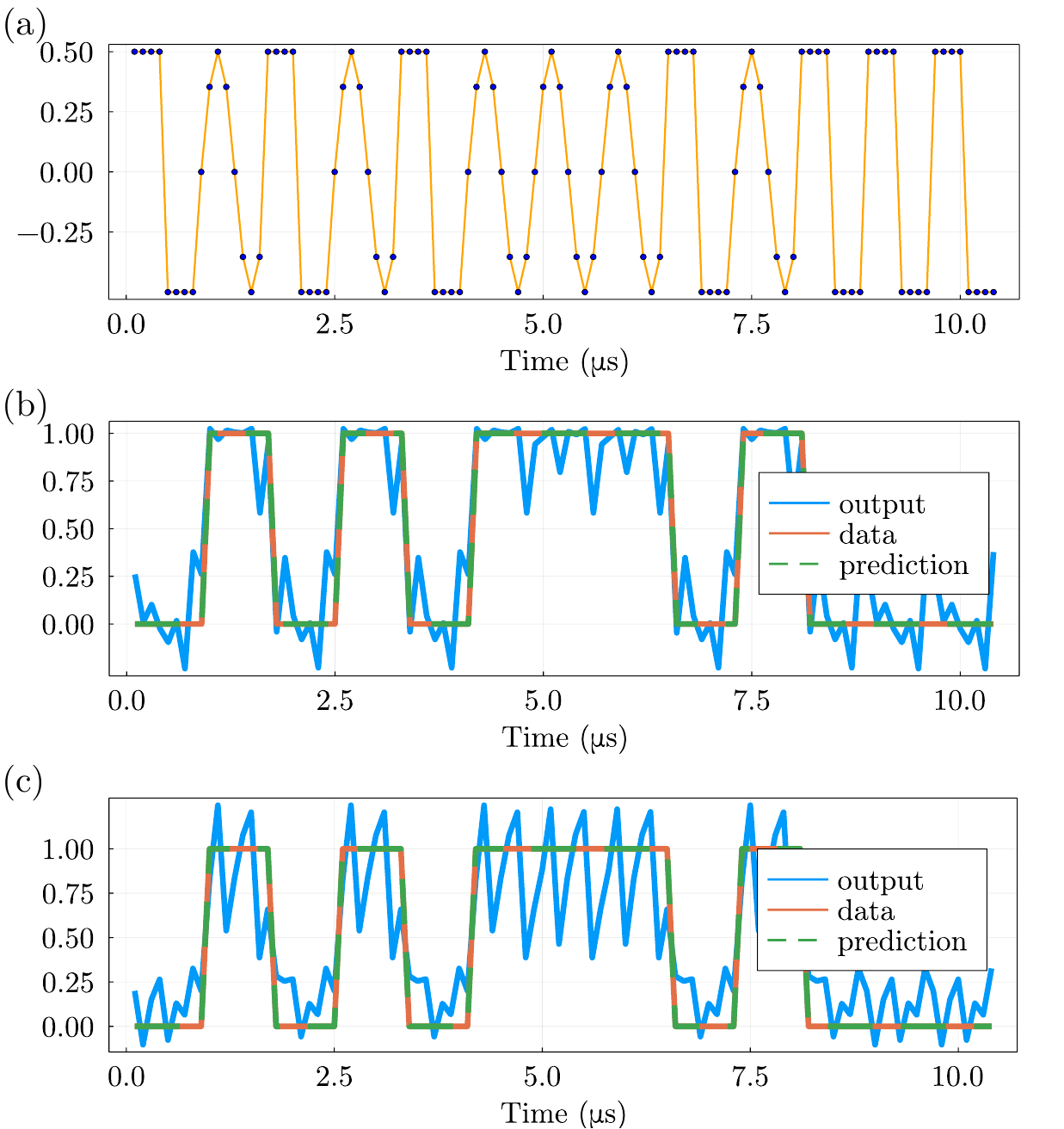}
\caption{(a) The input data for the sine and square waveform classification task is a time series of points belonging to a sine or a square discretized in 8 points. (b) Performance on the sine and square waveform classification task with 16 measured basis state occupations as neural outputs (up to $|33\rangle$) and (c) with 9 measured basis state occupations (up to {$|22\rangle$}). The target is shown in full orange line and the reservoir prediction in dashed green line. For simulations methodology see Methods.}
\label{sin_square}
\end{figure}

In order to evaluate the capacity of the quantum reservoir with oscillators, we address two standard benchmark tasks of reservoir computing, i.e. a classification task that requires a lot of nonlinearity and short-term memory, and a prediction task that requires both short- and long-term memory. In order to assess the advantage brought by the quantum nature of the reservoir, we compare its performance with that of classical reservoirs on the same tasks. We differentiate between the contributions of dynamic features and distinctively quantum properties, by conducting comparisons with both static and dynamic classical reservoirs. For the static reservoir we perform software simulations of reservoirs with neurons that apply a nonlinear ReLu function (typically used in machine learning). For the dynamical reservoir we simulate spin-torque nano-oscillators as neurons, such as they were used in \cite{RiouIEEE, Torrejon2017}, and also compare the different simulation results with the  experimental performance obtained in \cite{RiouIEEE}. The methodology and parameters used for simulations are described in Methods.

\subsection*{Sine and square waveform classification}

The first learning task that we address is the classification of points belonging to sine and square waveforms. The input data is sent as a time-series, consisting of 100 randomly arranged sine and square waveforms, each discretized in 8 points, as shown in FIG.~\ref{sin_square}(a). The neural network gives a binary output, equal to 0 if it estimates that the input point belongs to a square, and to 1 if it estimates that it belongs to a sine. This task was specifically conceived to test the nonlinearity and the memory of a neural network as the input data points cannot be linearly separated and the extremal points require memory to be distinguished (input points equal to $\pm 1$ can both belong to a sine and to a square). At least 24 classical neurons are needed to solve this task with an accuracy $>$99\%\cite{RiouIEEE}. 

\begin{figure}
\includegraphics[scale=0.35]{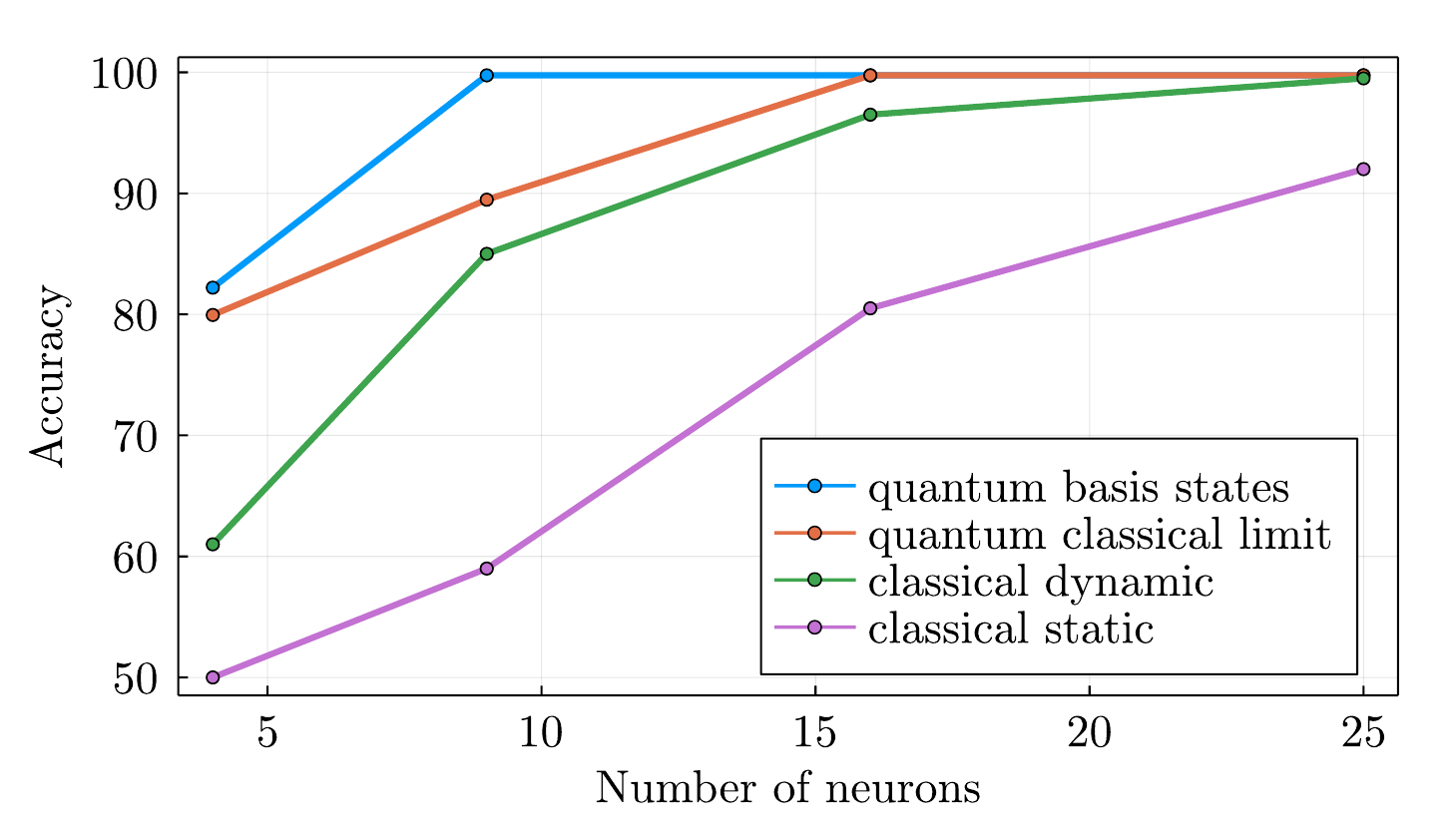}
\caption{Accuracy on the sine-square waveform classification task as a function of the number of measured neurons, for classical reservoir (both static and dynamic) and for quantum reservoir with quantum oscillators (including the classical limit obtained at large dephasing). For more details on simulations see Methods.}
\label{performance_neurons}
\end{figure}

\begin{figure*}
\begin{center}
\makebox[\textwidth][c]{\includegraphics[width=1.1\textwidth]{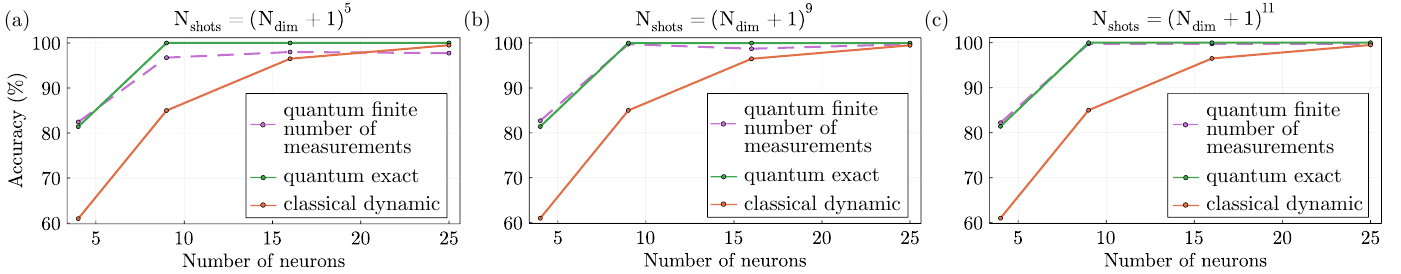}}%
\caption{Accuracy on the sine and square waveform task as a function of the number of measured neurons with a finite number of measurement samples $N_{\rm{shots}}$, compared to the ideal measurement of the quantum system (quantum exact) and classical dynamic. }
\label{shots}
\end{center}
\end{figure*}
 
We send the input drives to the oscillators for 100 ns, one immediately after the other, and we measure the occupation probabilities at the end of each drive. Half of the data is used for training, and another half for testing the performance. We investigate the performance of the quantum reservoir as a function of the number of measured basis states.
We first measure the states $|00\rangle$ to $|33\rangle$, yielding 16 output neurons. The reservoir prediction, obtained with Eq.~\eqref{prediction} is shown in FIG.~\ref{sin_square}(b). The prediction matches the target with 99.7 \% accuracy. This is a very good performance - indeed, it requires at least 40 static classical neurons and 24 dynamical classical neurons (see FIG.~\ref{performance_neurons} for simulations and ~\cite{RiouIEEE} for experiments) to achieve it. The fact that it is obtained with only 16 measured quantum neurons points to the first aspect of quantum advantage: all the 81 basis states that are populated participate to data processing and transformation even though they are not measured. We push this even further and perform learning while only measuring states up to $|22\rangle$, which yields 9 neurons. Strikingly, we still obtain an accuracy of 99~\% (FIG.~\ref{sin_square} (c)). Therefore, this task that requires at least 24 classical dynamical neurons, is perfectly solved by measuring only 9 quantum neurons.

To better understand where does this advantage come from, we also performed learning with our quantum system in the classical limit of large dephasing. In this limit, the oscillators exist in a statistical mixture of states instead of the quantum superposition, and quantum coherences vanish. Our observation shows that for 4 and 9 neurons, quantum oscillators perform better than in the classical limit (see FIG.~\ref{performance_neurons}), indicating that quantum coherences play a crucial role. Moreover, the classical limit outperforms classical spintronic oscillators, which can be attributed to the fact that even unmeasured basis states still participate in the transformation of input data. This aspect is interesting from an experimental perspective because, even though quantum measurements need to be repeated multiple times to reconstruct the probability amplitudes to find a system in a specific state, it implies that a much smaller number of states need to be measured in comparison to the classical case. Furthermore, all the measurements are performed on the same device, which simplifies the experimental setup, and enables simultaneous measurement using frequency multiplexing~\cite{Essig2021}.

Experimentally, the states would be measured by coupling a qubit to each resonator and using the dispersive readout. We study the number of measurements that are needed to obtain sufficiently precise basis state occupations in order to perform learning with the same accuracy as with the exact probability amplitude values~\cite{Khan2021}. The variance of the probability amplitude of the occupation of the states is given by the multinomial law
 \begin{equation}
 \langle p \rangle = \sqrt{\frac{p(1-p)}{N_{\rm{shots}}}},
 \end{equation}
 where $N_{\rm{shots}}$ is the number of measurements. For three different values of $N_{\rm{shots}}$, we add to the probability amplitude of the occupation of the states an error drawn from a Gaussian distribution of variance $\langle p \rangle$. Accuracy on the sine and square waveform classification task is shown in FIG.~\ref{shots}. For the small drives that we apply in our simulations, lower energy levels in each oscillator have higher probabilities to be occupied, which means that their measurement induces smaller errors. We find that for the first two levels in each oscillator (4 neurons in FIG. 5), we obtain with $(N_{\rm{dim}} + 1)^5$ shots sufficiently precise values to obtain the same accuracy as with the exact values. For 2 photon states (9 neurons in FIG. 5) with need $(N_{\rm{dim}} + 1)^9$ shots and for 3 and 4 photon states (16 and 25 neurons in FIG. 5) we need $(N_{\rm{dim}} + 1)^{11}$ shots.
 
Each measurement in a quantum system disrupts the coherence, making it necessary to remember prior inputs for time-dependent tasks. As a result, portions of the input sequence must be replayed before every measurement. However, recent studies have demonstrated that a quantum reservoir can effectively learn using weak measurements~\cite{Mujal2022a}. This approach eliminates the need for constant replays and subsequently reduces the overall duration of the experiment.

\subsection*{Mackey-Glass chaotic time-series prediction}

\begin{figure}
    \centering
    \includegraphics[scale = 0.45]{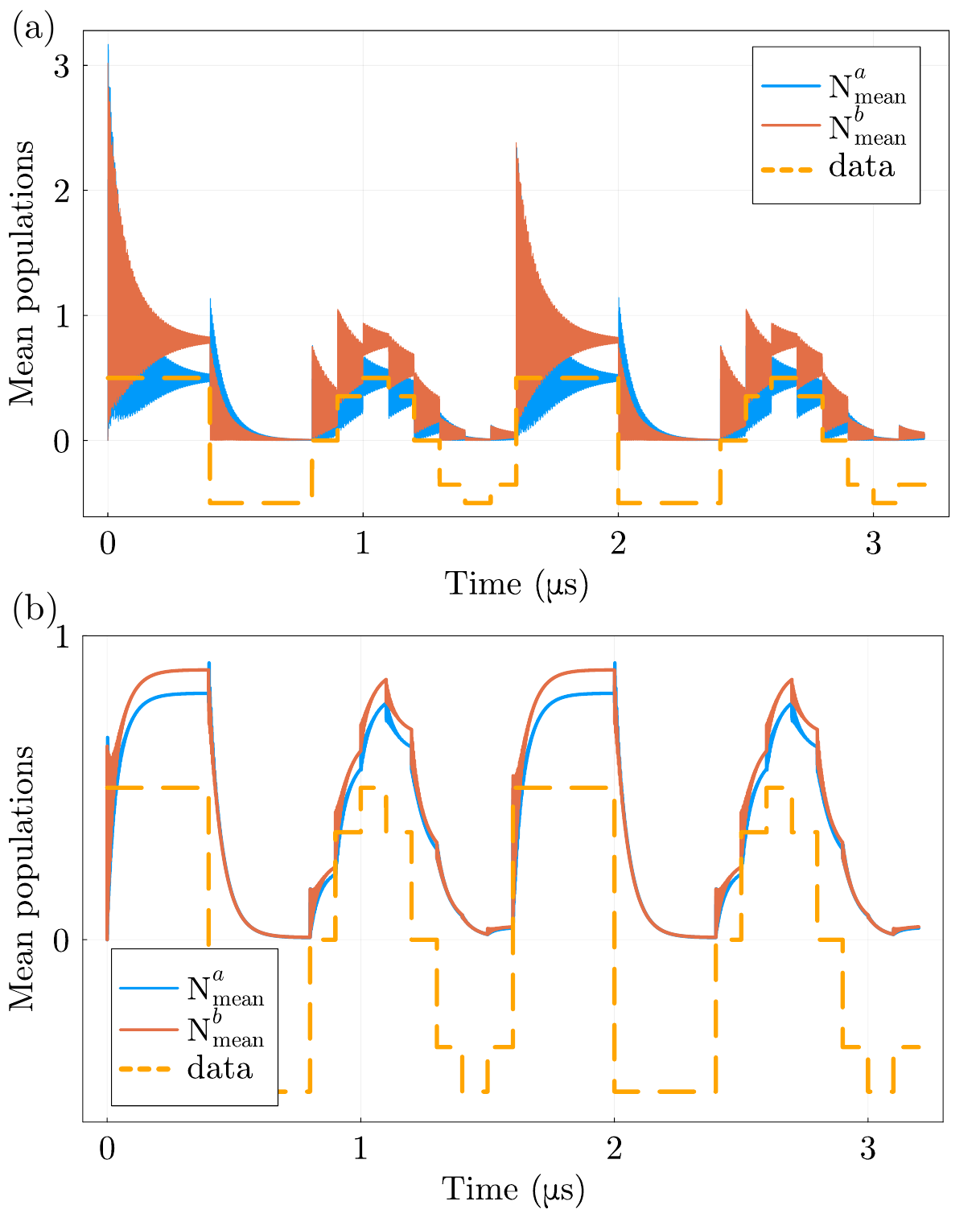}
    \caption{Comparison of mean photon populations in oscillators $N^a$ and $N^b$ for a sample input in the sine and square waveform classification task: (a) utilizing the quantum reservoir, and (b) in the classical limit featuring strong dephasing.}
    \label{dynamic_Na_Nb}
\end{figure}

The second benchmark task that we address is the prediction of Mackey-Glass chaotic time-series. Compared to classification, time-series prediction requires the reservoir to have an enhanced memory. It also allows us to investigate the impact of the reservoir temporal dynamics on its prediction capacity; in particular we study how the coupling between the oscillators and their dissipation rates impacts the reservoir performance. FIG.~\ref{dynamic_Na_Nb} shows that the dynamics of a quantum reservoir exhibits greater complexity than that of its classical limit. To make full use of this richness, we can sample the system occupations at several distinct times for a single input.

\begin{figure}
\includegraphics[scale=0.6]{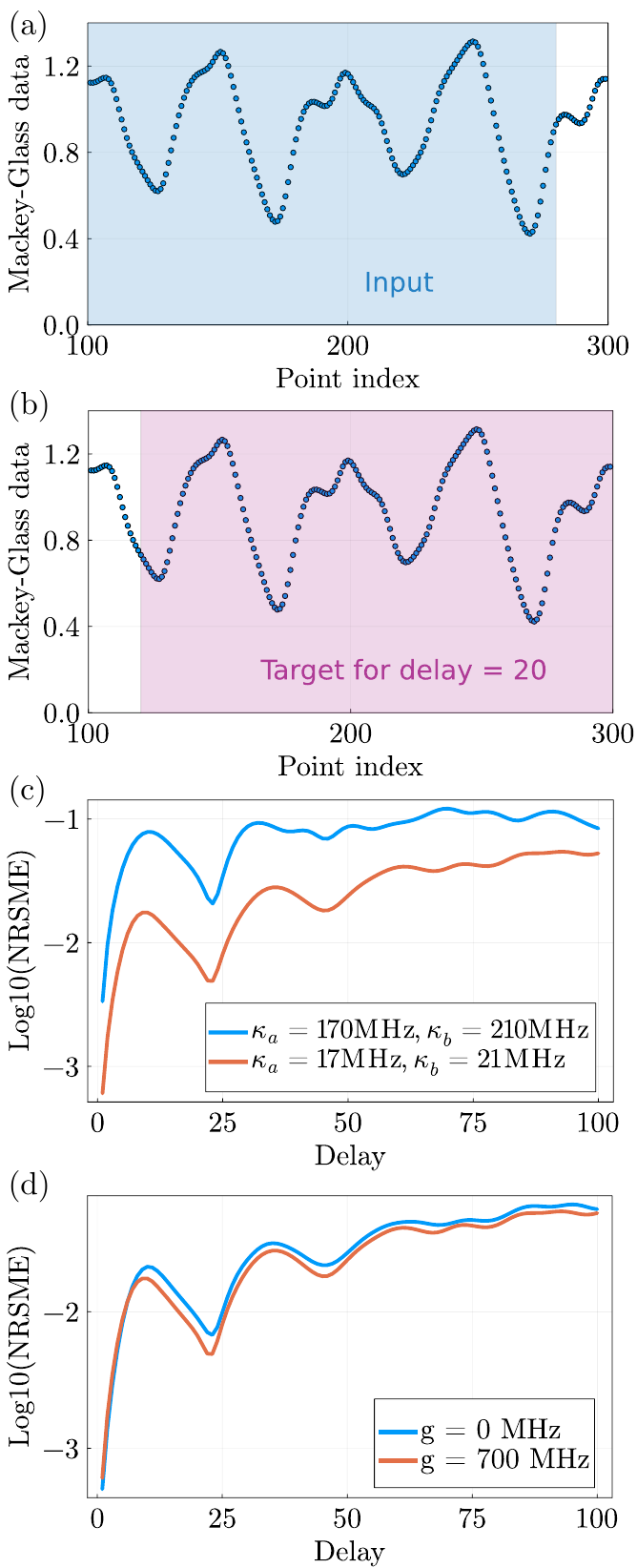}
\caption{(a) Mackey-Glass chaotic time-series data. For each input point, the target is the point with a certain delay in the future, here illustrated for delay=20.  (b-c) Logarithmic error of the reservoir for the Mackey-Glass task as a function of the number of points in future that the reservoir attempts to predict (b) for two different dissipation rates and for fixed coupling $g = 700$ MHz, and (c) for two different oscillator coupling $g$ values and fixed dissipation rates $\kappa_a = 17$ MHz and $\kappa_b = 21$ MHz.} 
\label{Mackey_Glass}
\end{figure}

The input data is obtained from the equation 
\begin{equation}
    \frac{\partial x(t)}{\partial t} = \frac{\beta x (t-\tau_M)}{1+x^{10} (t-\tau_M)} - \gamma x(t).
\end{equation}
It is chaotic for parameters $\beta=0.2$, $\gamma=0.1$ and $\tau_M = 17$ ~\cite{Chen2022}. A subset of the input data is shown in FIG.~\ref{Mackey_Glass} (a). The time here takes discrete values (point index). The task consists in predicting a point with a certain delay in the future. We have trained the reservoir for different delays varying from 1 to 100 (delay = 20 is shown in FIG.~\ref{Mackey_Glass} (a)). Each point is sent for 100 ns, such that delay = 20 corresponds to 2 $\mu$s. In all the simulations we measure 16 basis state neurons, from $|00\rangle$ to $|33\rangle$, and we sample the reservoir 10 times for each input, corresponding to a measurement every 10 ns.

We train the reservoir on 1000 points and test it on another set of previously unseen 1000 points. 
The results are shown in FIG.~\ref{Mackey_Glass} (b-c). We plot the average logarithmic error on 1000 test points as a function of the delay for different reservoir parameters such as dissipation rates and oscillator couplings. In all the cases, we observe an overall logarithmic increase of the error as a function of the delay, which corresponds to the memory of the reservoir - points further in future are harder to predict because the memory is lost. Still, the error saturates for large delays. This saturation is due to the fact that the reservoir learns the range in which the points are situated, and in particular the region where the minima and the maxima of the time-series, that contain a lot of points, are concentrated. Another common feature that can be noticed in all the figures are the oscillations in the error signal that reflect the periodicity in the input data.

We first study the impact of the oscillator dissipation rates $\kappa_a$ and $\kappa_b$ on the reservoir performance (FIG.~\ref{Mackey_Glass} (b)). We observe that for high dissipation, the error is globally larger, and most importantly, increases faster - meaning that the memory of the neural network is shorter. It is thus important to have high-quality-factor oscillators to solve tasks that require a lot of memory. Second, we study the impact of the oscillator coupling rate $g$ on the reservoir performance (FIG.~\ref{Mackey_Glass} (c)). For larger couplings the error decreases; indeed, strong coupling gives rise to multiple data transformations between different basis states which is essential for learning. It also leads to a significant population of a larger number of basis state neurons that contribute to computing.

In previous works, this task was solved in simulations with similar performance using 50 classical dynamical neurons such as skyrmions~\cite{Chen2022} and experimentally with a classical RC oscillator that was time-multiplexed 400 times to obtain 400 virtual neurons~\cite{Appeltant2011}. In comparison, here we solve it with just two physical devices and 16 measured basis state neurons.

\section*{Discussion}

We have shown that a simple superconducting circuit, composed of two coherently coupled quantum oscillators, can successfully implement quantum reservoir computing. This circuit has been exploited for quantum computing for years, and can be readily used to realize experimentally larger scale quantum reservoir computing.

Compared to classical reservoir, quantum reservoir allows to encode neurons as basis states and to obtain a number of neurons exponential in the number of physical devices. Furthermore, even though it was not the focus of this paper, where we processed classical data, numerical simulations of different quantum reservoirs have shown that quantum reservoirs can process input quantum states~\cite{Ghosh2019, Angelatos2020a} and simultaneously estimate their different properties, as well as perform quantum tomography~\cite{Ghosh2020}. This is particularly interesting in the age where quantum computing encodes information in quantum states and begins to produce more and more quantum data that will need to be automatically classified.

Quantum reservoirs can be implemented on different quantum systems~\cite{Fujii2017, Ghosh2019, Angelatos2020a, Nokkala2021, Govia2021a}. First works have naturally focused on qubits, as the most common quantum hardware~\cite{Fujii2017, Ghosh2019}. Nevertheless, quantum oscillators compared to qubits have a net advantage for scaling - they have an infinite number of basis states, compared to qubits that only have two - and they can be much more efficiently populated using resonant drives and coherent coupling. With just two quantum oscillators, we can populate up to 9 states in each oscillator with significant probability amplitudes, which yields a quantum reservoir with up to 81 neurons. By measuring only 16 basis states, we obtain a performance equivalent to 24 classical oscillators. There is an advantage in the number of physical devices, which simplifies experimental implementation, and also in the number of neurons that need to be measured, which simplifies the measurement procedure. 

 Reservoir computing was already simulated on a single nonlinear quantum oscillator~\cite{Kalfus2022} and on a system of coupled nonlinear parametric oscillators~\cite{Angelatos2020a}. Our work significantly differs from these approaches. In these works, quantum oscillators were operating in the semi-classical regime, where a strong input signal with a large number of photons induces Kerr nonlinearity. In that regime, each oscillator yields two output neurons, i.e. the two field quadratures, that can be sampled in time in order to increase the number of effective neurons. Our approach fully exploits the quantum nature of the system by using the basis states as neurons, which allows to increase the memory of the system as there is no need for sampling, and reduce both the number of physical devices in the system and the number of necessary measurements.

We believe that this solution is very promising for the implementation of quantum neural networks as it is scalable. Indeed, it has recently been shown that multiple oscillators can be parametrically coupled all-to-all through a common waveguide~\cite{Zhou2021}. This new paradigm that we propose would thus allow to realize larger scale quantum neural networks with readily available devices.

\section*{Methods}

\subsection{Quantum simulations}

We simulate the dynamics of the coupled quantum oscillators using the library QuantumOptics.jl for simulating open quantum systems in Julia~\cite{Kramer2018}. The dynamics can be captured by the quantum master equation
\begin{equation}
    \dot{\rho} = -i [\hat{H} + \hat{H}_{drive}, \rho] + \hat{C}\rho \hat{C}^\dagger - \frac{1}{2} \hat{C}^\dagger \hat{C} \rho - \frac{1}{2} \rho \hat{C}^\dagger \hat{C}, 
\label{master_equation}
\end{equation}
where $\rho$ is the density matrix of the system. $\hat{H}_{drive}$ is the resonant drive Hamiltonian~\cite{Gardiner1993, Carmichael1993, Kiilerich2019a} 
\begin{equation}
    \hat{H}_{drive} = i \epsilon_a \sqrt{2 \kappa_a} (\hat{a} - \hat{a}^\dagger) + i \epsilon_b \sqrt{2 \kappa_b} (\hat{b} - \hat{b}^\dagger)
\end{equation}
and
\begin{equation}
\epsilon_a = \epsilon_b = \epsilon_0 \times x_i
\label{input_renormalization}
\end{equation}
are the drive amplitudes that encode the input data $x_i$. For sine-square waveform classification task we use $\epsilon_0^a = \epsilon_0^b = 1.2 \times 10^6 \sqrt{\rm{Hz}}$ in order to populate with a significant probability the first 5 levels in each oscillator, and to have negligible probability to populate states above 8 (we truncate the Hilbert space at 8 photons in each oscillator).
The Mackey-Glass data takes, on average, larger values compared to sine and square waveform classification - we thus use a smaller value for the drives amplitudes, $\epsilon_0^a = \epsilon_0^b = 5 \times 10^5 \sqrt{\rm{Hz}}$.

We consider that oscillators couplings to transmission lines $\kappa_a$ and $\kappa_b$ are dominant terms in the oscillators dissipation, such that we can neglect the internal losses. The collapse operator associated with the decay in the modes $a$ and $b$ can thus be written as 
\begin{equation}
\hat{C} = \sqrt{\kappa_a} \hat{a} +  \sqrt{\kappa_b} \hat{b}.
\end{equation}
Finally, reservoir outputs are obtained as the expectation values of the basis states occupations
\begin{equation}
   p(n_a, n_b) = \langle n_a n_b | \rho | n_a n_b \rangle.
\end{equation}

\subsection{Classical limit of the quantum system}

We simulate the classical limit of our quantum reservoir by adding a large dephasing term to the collapse operator in the matrix equation
\begin{equation}
\hat{C} = \sqrt{\kappa_a} \hat{a} +  \sqrt{\kappa_b} \hat{b} + \sqrt{\kappa_{\phi}} (\hat{n}_a + \hat{n}_b),
\end{equation}
with $\kappa_{\phi} = 100$ MHz.
With dephasing, the mean values of the photon numbers in oscillators are higher than without dephasing, and we would need to increase the size of the simulated Hilbert space to $N_{\rm{dim}}$=12, which would make the simulations too computationally demanding. This is why in the simulations with dephasing we decrease the input drive amplitudes to $\epsilon_a^0 = \epsilon_b^0 = 5.5 \times 10^5 \sqrt{\rm{Hz}}$, which gives the same mean number of photons as for simulations without dephasing.

\subsection{Classical static reservoir simulations}

The simulations of the classical reservoirs, both static and dynamic, were performed in the library \textit{pytorch} for training neural networks in Python. 

The state of the reservoir at time $t$ is
\begin{equation}
   y(t) = f( W_{\textrm{in}} x(t) + W_{\textrm{res}} y(t-1) )
\end{equation}
where $f$ is the ReLu function, $W_{\textrm{in}}$ is the vector that has the length of the size of the reservoir, and maps the input data on the reservoir, $W_{\textrm{res}}$ is a square matrix that has the dimension of the size of the reservoir and which gives the memory to the reservoir. Here the reservoir has the memory of a single step in the time, which is sufficient for the sine and square waveform classification task. In the simulations of the static reservoir, the size of the reservoir is equal to the number of measured neurons, shown in FIG.~\ref{performance_neurons}.

\subsection{Classical dynamic reservoir simulations}

The simulations of dynamical classical reservoir were realized considering a spin-torque nano-oscillator as a neuron, as in~\cite{Torrejon2017}. The dynamics of the nano-oscillator can be modeled as that of a nonlinear auto-oscillator~\cite{Slavin2009}
\begin{equation}
    \frac{dp}{dt} = 2(-\Gamma (1+Qp) +W_{\textrm{in}} I \sigma (1-p)) p
    \label{Slavin}
\end{equation}
where $p$ is the power of the oscillator, $\Gamma$ is the damping rate, $Q$ is the non-linearity, $I$ is the current that drives the oscillator and $\sigma$ is a factor related to the geometry of the oscillator.
Input data is encoded in the current $I$ and mapped by the randomly generated vector $W_{\textrm{in}}$ on the reservoir. Reservoir outputs are obtained from the oscillator power $p$ by numerically integrating the Eq.~\eqref{Slavin}.

\section*{Data availability}

Data and code are available on Zenodo: https://doi.org/10.5281/zenodo.7817435.

\section*{References}
\nocite{*}
\bibliography{quantum_res.bib}

\section*{Acknowledgements}

This research was supported by the Quantum Materials for Energy Efficient Neuromorphic Computing (Q-MEEN-C), an Energy Frontier Research Center funded by the U.S. Department of Energy (DOE), Office of Science, Basic Energy Sciences (BES), under Award DE-SC0019273 and by the Paris Île-de-France Region in the framework of DIM SIRTEQ.

\section*{Author contributions}

D.M. and J.G. conceived the project. D.M. performed the calculations. J.D performed the quantum simulations. B.C. participated to analytical calculations and quantum simulations. E.P. and A.M. performed the classical dynamical simulations. D.M. performed the classical static simulations. D.M. and J.G. wrote the paper.

\section*{Competing interests}

The Authors declare no Competing Financial or Non-Financial Interests.

\end{document}